\documentclass[12pt, letterpaper]{article}

\usepackage{graphicx}
\usepackage{todonotes}
\usepackage{xurl}

\AtBeginDocument{%
  \providecommand\BibTeX{{%
    \normalfont B\kern-0.5em{\scshape i\kern-0.25em b}\kern-0.8em\TeX}}}

\usepackage[letterpaper, left=1in, right=1in, bottom=1in, top=0.75in]{geometry}

\author{
Nikita Belenkov\\
\textit{Quantstamp, Inc.}\\
nikita@quantstamp.com
\and
Valerian Callens\\
\textit{Quantstamp, Inc.}\\
valerian@quantstamp.com
\and
Alexandr Murashkin\\
\textit{Quantstamp, Inc.}\\
alex@quantstamp.com
\and
Kacper Bak\\
\textit{Quantstamp, Inc.}\\
kacper@quantstamp.com
\and
Martin Derka\\
\textit{Zircuit}\\
martin@zircuit.com
\and
Jan Gorzny\\
\textit{Zircuit}\\
jan@zircuit.com
\and
Sung-Shine Lee\\
\textit{Quantstamp, Inc.}\\
martinet@quantstamp.com
}

\begin{document}

\title{SoK: A Review of Cross-Chain Bridge Hacks in 2023}









\date{19 December 2023}

\maketitle


\begin{abstract}

Blockchain technology has revolutionized industries by enabling secure and decentralized transactions. However, the isolated nature of blockchain ecosystems hinders the seamless transfer of digital assets across different chains. 
Cross-chain bridges have emerged as vital web3 infrastructure to address this challenge by facilitating interoperability between distinct blockchains.
Cross-chain bridges remain vulnerable to various attacks despite sophisticated designs and security measures. The industry has experienced a surge in bridge attacks, resulting in significant financial losses. The largest hack impacted Axie Infinity Ronin Bridge, with a loss of almost \$600 million USD.
This paper analyzes recent cross-chain bridge hacks in 2022 and 2023 and examines the exploited vulnerabilities. By understanding the attack nature and underlying weaknesses, the paper aims to enhance bridge security and propose potential countermeasures. The findings contribute to developing industry-wide standards for bridge security and operational resilience. Addressing the vulnerabilities and weaknesses exploited in recent cross-chain bridge hacks fosters trust and confidence in cross-chain interoperability.

\end{abstract}

\sloppy
\section{Introduction}\label{sec:intro}

In recent years, the rapid growth of blockchain technology has revolutionized various industries, enabling secure and decentralized transactions. 
One of the significant challenges blockchain networks face is the isolated nature of their respective ecosystems, hindering the seamless transfer of digital assets across different chains. 
The industry has entered a multi-chain era, with countless new chains following different technological approaches. 
One common one is the Ethereum network which uses a dedicated virtual machine, the Ethereum Virtual Machine (``EVM'') \cite{yellow_paper}, to support the execution of code, called \emph{smart contracts}. 
Improving the interoperability of these chains became a priority to enable seamless interaction across chains while preserving the tenets of the native chains. 
This allows for the reduction of fragmentation of the ecosystem, an increase in the utilization of liquidity, and enables scalability.

Cross-chain bridges serve as vital infrastructure components, enabling the transfer of digital assets and data between distinct blockchain ecosystems. These bridges rely on complex system architectures involving multiple components, including custodians, communicators, and debt-issuers, to ensure the secure and reliable transfer of assets. 
However, despite these bridges' sophisticated design and security measures, they remain susceptible to various vulnerabilities and attacks. 
The blockchain industry has made a lot of progress toward a more secure ecosystem, but in recent years, bridges have been the main target for attacks. 
Nearly \$2 billion USD have been stolen in bridge hacks \cite{Zach_2022} since 2020, with the Axie Infinity Ronin Bridge being the most significant hack: a loss of \$600 million USD \cite{ronin-rekt}. 
For context, the total value locked in different types of token bridges reached \$10 billion USD and, at the time of writing this paper, stood at around \$5.5 billion USD\footnote{Data from L2BEAT (https://l2beat.com/scaling/tvl) as of 27 Nov 2023.}. 
 
There are two main reasons why bridges are often attacked.
First, a common design of cross-chain bridges consists of locking a significant amount of tokens in only one or two contracts, creating a very high reward if the attack is a success. 
Second, these systems usually have a much larger attack surface than ordinary blockchain decentralized applications (``dApps'').
This is because they have on-chain components on two separate blockchains, as well as off-chain components to communicate between these blockchains. 
Multiple hacks happened due to poor information technology practices rather than issues with on-chain code.

The goal of this paper is to analyze recent cross-chain bridge hacks that happened after \cite{prev_paper_QS} and provide an in-depth examination of the vulnerabilities exploited. 
By understanding the nature of these attacks and the underlying weaknesses they exploit, we can enhance the security of cross-chain bridges, propose potential countermeasures to mitigate such risks, and inform the future of industry-wide standards for development and monitoring.

The paper is structured as follows: in the next section, we delve into the architectural aspects of cross-chain bridges, outlining the key components (custodian, communicator, debt issuer) and their functions. This foundational understanding is crucial to describe the attack vectors in the subsequent sections. We then examine custodian attacks in Section \ref{sec:custodian}. Next, we explore communicator attacks in Section \ref{sec:communicator}, which exploit vulnerabilities associated with the communicator of cross-chain bridges. These attacks primarily focus on intercepting and manipulating or falsifying the data transmitted between the connected networks. By analyzing recent instances of these attacks, we can identify common patterns and potential security weaknesses in bridge communication protocols.
Following the analysis of these attack vectors, we discuss related work in Section \ref{sec:related} that has contributed to better understand cross-chain bridges. Additionally, we highlight future research directions and potential solutions to enhance the security of cross-chain bridges and prevent similar attacks. 
Finally, we summarize the key findings from the analysis of recent cross-chain bridge hacks in Section \ref{sec:conclusion}.

\section{Bridge Architecture}\label{sec:bridges}

In this section, we describe the general architecture of a cross-chain bridge. The term ``cross-chain'' suggests the existence of two blockchains, or ``chains''. For simplicity, we limit our analysis to chains supporting the execution of smart contracts. 
The bridge is used to transfer a digital asset from a source chain to a destination chain according to different strategies, depending on where that asset can be created or ``minted''. 
In the rest of this document, we will only consider the simple case where the asset can only be created on the source chain. The digital asset can be a native asset (for instance, Ether on Ethereum) or an asset represented by a smart contract (for instance, a contract implementing the standard ERC-20 \cite{ERC-20} to represent fungible tokens or a contract implementing the standard ERC-721 \cite{ERC-721} or ERC-1155 \cite{ERC-1155} to represent non-fungible tokens).

For a cross-chain transfer to happen, the first step is to create a smart contract on the destination chain that would correspond to the token and contract on the source chain. 
Anyone owning the token on the source chain could lock that asset in the source chain smart contract, and the corresponding smart contract on the destination chain would unlock (or mint) a corresponding token.
We call that contract the ``debt issuer'' and the token the ``debt token''.
Once a digital asset has a corresponding contract on the destination chain owned by the bridge authority, the bridge authority must deploy a custodian contract on the source chain where the asset can be deposited by any user who wants to perform a cross-chain transfer. 
Consistent rules ensure that it is only possible to unlock these assets from the custodian by meeting some conditions, such as receiving a valid cross-chain transfer from the destination chain. 
The next two paragraphs describe what happens in each direction: when tokens are locked in the custodian and debt tokens are minted by the debt issuer, and when debt tokens are destroyed by the debt issuer and tokens are unlocked by the custodian. 

\begin{figure}
    \centering
    \includegraphics[scale=0.6]{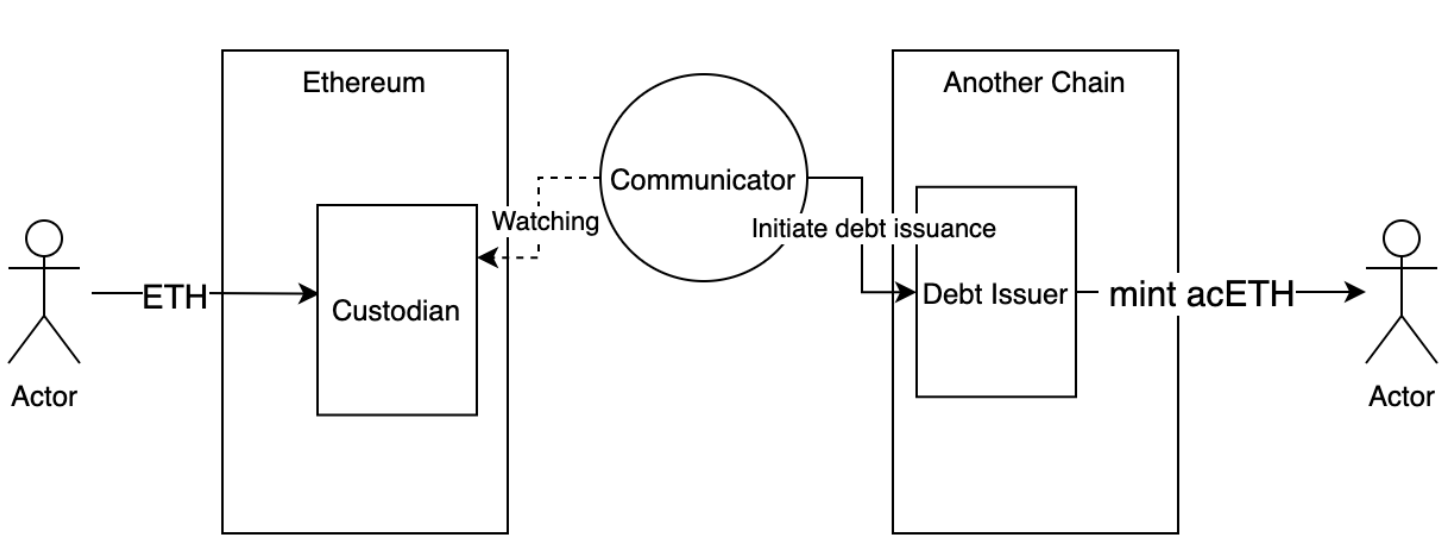}
    \caption{Cross-chain transfer from a source chain (Ethereum) to a destination chain (Another Chain). The digital asset is deposited (ETH) into the custodian contract, which triggers a reaction by the Communicator that will interact with the debt issuer contract to mint debt tokens (acETH) in favor of the cross-chain transfer recipient.}
    \label{fig:architecture-flow-source-to-destination}
\end{figure}

When a valid deposit has been recorded in the custodian on the source chain, a transaction must be executed on the debt issuer deployed on the destination chain. For that, someone has to play the role of a communicator, which consists of three main tasks: watching the custodian contract to detect deposit requests, confirming the validity of the cross-chain transfer, and relaying the request by interacting with the debt issuer contract. Some tasks can be executed automatically or manually by the same or different actors. 
An example is to have a custodian emitting a specific \texttt{event}\footnote{An \texttt{event} can be considered a log generated by a smart contract during its execution in a format that can easily be monitored by external observers and that can be used as a way to asynchronously trigger off-chain actions.} when a deposit is accepted. Then, a centralized actor monitors these events emitted by the custodian contract. It confirms the event's validity by waiting a sufficient amount of time to ensure that the nodes securing the source chain agreed that it was a valid request. Finally, it has the authority to mint debt tokens on the destination chain in favor of the recipient of the cross-chain transfer. The whole process is illustrated in Figure \ref{fig:architecture-flow-source-to-destination}.

\begin{figure}
    \centering
    \includegraphics[scale=0.6]{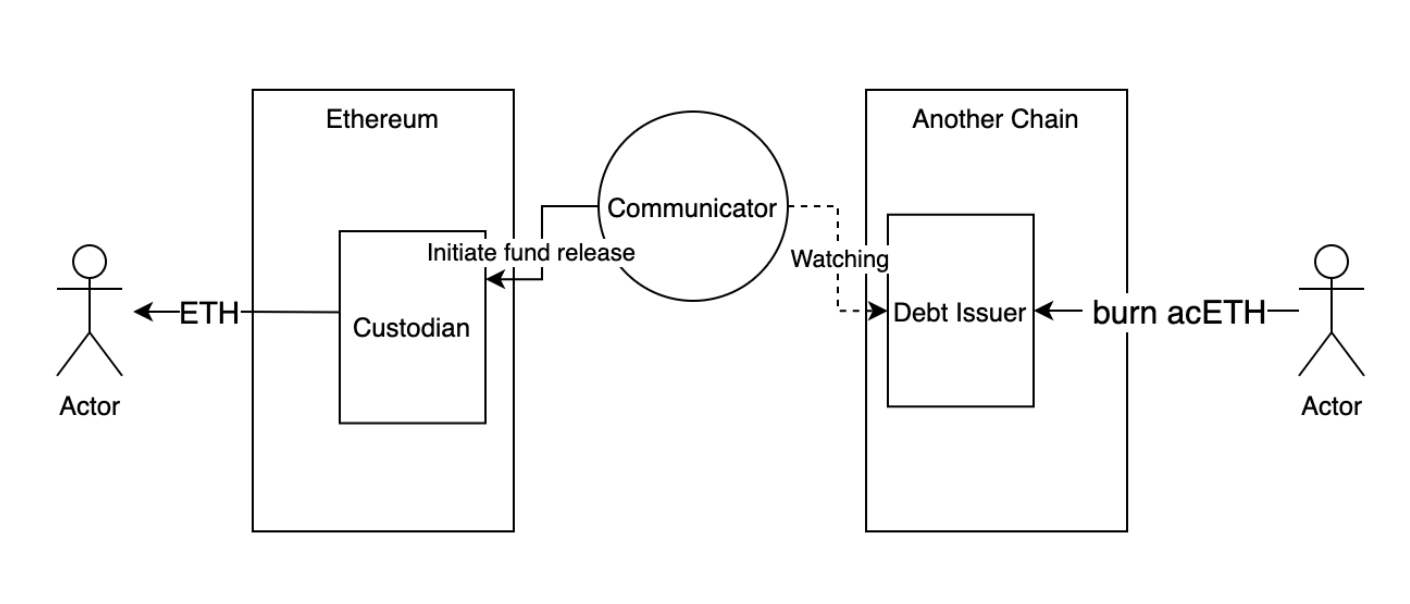}
    \caption{Cross-chain transfer from a source chain (Another Chain) to a destination chain (Ethereum). The debt tokens (acETH) are burned in the debt issuer contract, which triggers a reaction by the communicator that will interact with the custodian contract to unlock the digital asset (ETH) in favor of the cross-chain transfer recipient.}
    \label{fig:architecture-flow-destination-to-source}
\end{figure}

On the debt issuer contract side, any owner of debt tokens can initiate a cross-chain transfer from the debt issuer contract to the custodian contract. For that, the debt issuer contract has to burn (or destroy) these tokens. Then, once again, a communicator must monitor the debt issuer contract, confirm the validity of the request, and relay the request by interacting with the custodian contract to unlock the tokens in favor of the recipient of the cross-chain transfer. The whole process is illustrated in Figure \ref{fig:architecture-flow-destination-to-source}.

Several security aspects must be considered to limit the risks faced by this bridge architecture. The attack surface of a bridge is a combination of the attack surface of each of its components and where the current risk exposure depends on the weakest components. In terms of integrity, clear and consistent rules must be defined for each component. Modifying the internal ledger storing the asset balances should not be possible for external actors. Regarding access control, a valid signature of the communicator must be mandatory to unlock assets from the custodian and mint debt tokens on the debt issuer. If not, the one-to-one peg between original and debt tokens would not be guaranteed.

Another risk is related to the communicator relaying an incorrect message because it did not wait for the message to be final on the source chain. Such a message still needs to be confirmed by the nodes securing the source chain. Regarding availability, the components should have a liveness guarantee to ensure that each cross-chain request will always be executed on the destination chain or aborted on the source chain to prevent requests from remaining forever in a pending state. 
This liveness guarantee has a cost for the communicator. 
He has to maintain an infrastructure able to constantly monitor the source chain and relay messages to the destination chain. 
As a result, the communicator should be positively incentivized to act in favor of the system and negatively incentivized to act against the system. If not, there may be occasions where the communicator could conclude that it is more profitable to act against the system.

After outlining general bridge architecture and potential risk aspects, it is essential to highlight that no bridges are the same. The implementation aspect differs highly between different bridges. However, there are two main types of bridges: trusted and trustless. These terms mainly describe how the interaction occurs between the two chains and which types of entities can act as communicators. The following two sections will cover the general ideas behind the two types and specific examples of implementations of those theoretical ideas. 

\subsection{Trusted bridges}

Trusted bridges are one of the common designs of bridges, where the role of the communicator is assigned to specific actors by the bridge team.
They do not require complicated proofs to be submitted across contracts. 
Instead, simple access control is necessary, where only a specific group of actors have the elevated permission to submit actions. 
The next three paragraphs present three types of trusted bridges.

The first type is a single entity that relays messages across chains, where both sides of the chain fully trust this entity.
Only this entity can interact with the bridge contracts. 
Centralized exchanges use this common design to move their assets across chains. 
This introduces apparent issues where the bridge's owner has absolute control over the assets and, if compromised, can lead to catastrophic events (e.g., stealing user's funds or minting tokens out of thin air).

The second type is a set of entities that validate the transfer of messages across chains. 
These entities can be called \emph{validators}. 
The design is close to the previous type because a limited number of actors can impact the behavior of the bridge contracts. The process of becoming a validator varies from project to project. Some projects manage all validators internally, while others allow external parties to manage them. The number of validators also varies significantly. The extra protection comes from the fact that a consensus about messages to relay across chains must be reached by the set of validators, making it more complicated and costly to trigger a malicious behavior on the destination chain. An example is the Ronin Bridge described in the next section.

The third type are so-called \emph{optimistic} bridges. 
The idea here is to build a system that will work correctly if at least one actor remains honest. 
Optimistic bridges introduce an observer party concept that ensures that the bridge behaves correctly and can react when malicious activity occurs.
This can be achieved by adding a mechanism that requires users to stake funds that can be slashed if they are proven to be fraudulent.
For example, if a validator communicates some incorrect data across the bridge boundaries, then the observer would detect this and craft proof that fraud has occurred. 
Once this proof is verified by the bridge contracts, the malicious validator will get slashed, meaning that it will get punished for bad behavior financially (or in some other way). 
%
Hence optimistic bridges require only one honest observer to behave correctly, as they can watch all other actors. 
This system makes semi-centralized communicators be passively observed, incentives good behavior and punishes (or mitigates) bad ones. 
There are limitations to this approach, like the complexity of handling rollbacks if malicious behavior is detected. 
These bridges work best when anyone can become an observer, so that honest users can become this actor themselves if necessary.

\subsection{Trustless bridges}

A logical continuation of trusted bridges is to remove the trusted communicator. This results in making the communicator network permissionless to join, so anyone could become a communicator.

One of the ways to design such a trustless bridge is called a \emph{state validating bridge}. 
In the case of state validating bridges or trustless bridges in general, state or changes of state of the blockchain are now transferred across both chains rather than only specific events like in the trusted approach. 
Once the communicator has relayed the new block from the source network, the state validating bridge validates the whole block to make sure it is correct and follows all the rules, and conditionally updates the internal state of the system. 
One common use case of this design is a so-called rollup (also known as a \emph{commit-chain} \cite{cryptoeprint:2018/642}) where the bridge also posts the rollup chain data, like its state root (see, e.g.~\cite{idealescapehatches}). 

Another implementation of the previous system is a \emph{consensus validating bridge}. 
This type of bridge also relays block headers across both chains, but does not verify the validity of transactions.
The destination chain receiving the block headers assumes that the source chain generating the headers came to consensus on which headers contain valid blocks and transactions.
In such a case, the destination chain may use a so-called \emph{light client} to check which source chain headers are valid.
A light client is a software application that interacts with the (source) blockchain network but does not store the blockchain. 
This application simply queries other nodes for the specific transactions and blocks in which it is interested.


Another recent variation of a trustless bridge is a \emph{zero-knowledge consensus verifying bridge}. 
Here, the state updates on the source chain are computed using a zero-knowledge system, which generates the updated state as well as a cryptographic proof that it was computed correctly (\cite{Groth16} is an example of such a proof system).
The phrase \emph{zero-knowledge} is used because these systems implement a method by which one party (the prover) can prove to another party (the verifier) that a given statement is correct while the prover avoids conveying any other information than the correctness of the statement.
However, these systems are not always used for privacy: the proofs are also \emph{succinct}, meaning that they quicker to verify than running the computation itself, and this property makes these systems efficient.
This is a recent research area being actively explored \cite{zk_bridge}. The idea here is to outsource the light client state verification of the previous type to an off-chain component, which would make a zero-knowledge proof of the state that would be checked on-chain. The main idea is that it is very difficult to build these consensus verifying smart contracts, which leads to high costs and security vulnerabilities. Having the consensus verification off-chain would make these bridges simpler and more viable. 
This research is still in development but is showing promising results \cite{Telepathy_2023}.

\section{Custodian Attacks}\label{sec:custodian}

In this section, we review four exploits that have exploited the custodian component of bridges. 
The first exploit involves manipulating a Merkle tree proof to add a non-existent deposit to the custodian contract.
The second exploit bypasses the validation of the transaction via a missing input validation on default values of transactions that have been added during an upgrade.
The third exploit uses compromised private keys to steal funds from the custodian.
Finally, the fourth exploit uses missing input validation in the legacy codebase to emit fraudulent deposit events.

\subsection{Incorrect Merkle proof validation and proof manipulation}\label{sec:cust-attack-1}
\begin{figure*}
    \centering
    \includegraphics[scale=0.7]{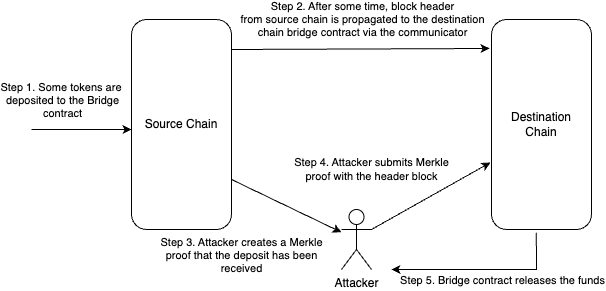}
    \caption{Overview of the normal interaction with the bridge using Merkle proofs.}
    \label{fig:binance-general}
\end{figure*}

One of the common decentralized bridge designs is a system where the headers of blocks are synchronized across both chains so that the transactions inside those blocks can be verified and executed correctly. 

In such a system, a user would provide proof of their deposit from one end of the bridge to the other so that the tokens can be minted or released on the other side. A general flow of this interaction can be seen in Figure \ref{fig:binance-general}. The following steps can be observed in the flow:

\begin{enumerate}
    \item The user deposits funds in the source chain bridge contract. This leads to a deposit event being emitted.
    \item After some time, the newly created block on the source chain will be propagated via the special communicator nodes to the destination chain. 
    \item To prove that a deposit exists on the source chain, the user (acting as a communicator) crafts a Merkle proof of the deposit, which would be represented as a leaf in the tree. 
    \item The user then submits that Merkle proof along with the block header number to prove to the source bridge contract that they have indeed deposited funds.
    \item The destination bridge contract releases or mints the funds to the user.
\end{enumerate}

\begin{figure}
    \centering
    \includegraphics[scale=0.5]{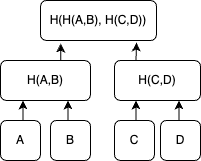}
    \caption{An example of a Merkle Tree. The leaf nodes are \texttt{A}, \texttt{B}, \texttt{C}, and \texttt{D}, and the parent node of \texttt{A} and \texttt{B} has the value of the hash of \texttt{A} and \texttt{B}, noted \texttt{H(A,B)}. Hence the root of the tree is the hash of its two child nodes, meaning \texttt{H(H(A,B), H(C,D))}.}
    \label{fig:merkle-tree}
\end{figure}

The key reliance of the system here is on these Merkle proofs. 
Different chains use different variations of these proofs that have their upsides and downsides. 
Figure \ref{fig:merkle-tree} shows an example of such a Merkle Tree. 
These trees rely on hashes that are one-way mathematical primitives that change if the input to the hash changes. It is easy to compute a hash of something but computationally infeasible to compute the input to a hash function, given a computed hash \cite{hash_func}.

Using a Merkle Tree as shown in Figure \ref{fig:merkle-tree}, it is possible to prove that there is a leaf \texttt{A} in the tree given the root hash and the path to the leaf. 
Here leaf \texttt{A} would represent a transaction, and the whole tree would record transactions in a block. 
Hence, propagating the root hash could prove that a transaction is a part of this state. 
The other content of the root hash also differs from blockchain to blockchain but usually includes the previous root hash to prove that the block is a child of the previous block.

There are multiple attack vectors in such a system. For example, one could consider manipulating such Merkle proof to accept fraudulent transactions. This would allow receiving funds from the destination chain without locking funds on the source chain. This would involve in the description above manipulating a proof in step 3 and then submitting a fraudulent one in step 4. 

\subsubsection{Real World Example}

\begin{figure*}
    \centering
    \includegraphics[scale=0.6]{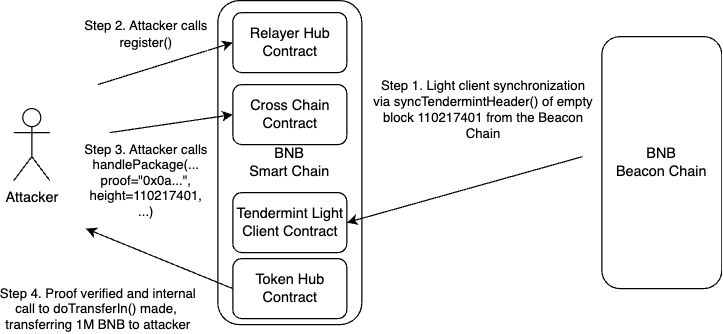}
    \caption{Overview of the attack of the Binance Bridge.}
    \label{fig:binance-diagram}
\end{figure*}

In this section, we review an attack that occurred on Binance Bridge in October 2022 and resulted in the loss of almost \$600 million USD \cite{bnb_rekt, samczsun-bnb, Igambediev-bnb}. 
Figure \ref{fig:binance-diagram} showcases the steps to exploit the Binance Bridge. Step 1 is a prerequisite for the attack: the \texttt{syncTendermitHeader()} function was used to sync the header of block number \texttt{110217401} from the Beacon Chain, dated August 29, 2020. This block has unusual features: it is empty (i.e., it contains no transactions) and the \texttt{proof} parameter is much shorter than the proofs passed to legitimate transactions. 

Step 2 of the attack to gain access was to register as a relayer by paying a fee of 100 BNB. 
Relayers possess the necessary access rights to interact with the BSC: Cross Chain contract, which is crucial for manipulating token transfers. The actual attack begins in Step 3 by calling the \texttt{handlePackage()} function of the BSC: Cross Chain contract and the following steps occur:
\begin{enumerate}
    \item As input to the \texttt{handlePackage()} function, the attacker provides the block number of \texttt{110217401} and a specially crafted proof. Then, the \texttt{handlePackage()} function validates the Merkle proof in function \texttt{validateMerkleProof()} that relies on the precompiled contract of the Binance Smart Chain at address \texttt{0x65}. 
    The precompiled contract uses the \texttt{iavlMerkleProofValidate()} function to run the actual validation. 
    \item The precompiled contract was a fork of the Cosmos Cross Chain Bridge protocol and contained a Merkle Tree verification function bug. 
    Binance uses a special type of Merkle Tree called an IAVL tree, which is a type of binary search tree that can only have one child per node \cite{bnb-iavl}. 
    IAVL trees are balanced binary search trees.
    \item The Merkle proof was manipulated as follows.
    A new leaf was cleverly inserted into the tree as a right leaf, along with the necessary (empty) internal nodes, in such a way that the tree was balanced and considered valid by the verification function.
    Due to the bug in the internal IAVL tree parsing, this change did not affect the root hash: the bug was that the library did not expect internal nodes to have both left and right children.
    So the new right leaf was not used in root hash computation and the transaction was successfully verified. 
    \item The extra leaf introduced a malicious 1M BNB deposit transaction that was then parsed and successfully executed by a call to the \texttt{doTransferIn()} function of the BSC: Token Hub contract.
\end{enumerate}
Cosmos and Binance have since issued a fix to reject any transaction with an IAVL tree with both left and right leaves.

\subsubsection{Solution}

This attack could have been mitigated by countering step 3 of Figure \ref{fig:binance-diagram} as this is the step in which the bug in the parsing was exploited. 
There are multiple approaches to attempt to mitigate the risk of a similar attack:

\begin{enumerate}
    \item Verify all third party libraries that the codebase is using, as here the bug was introduced via forking the Cosmos’ cross bridge framework.
    \item Verify all the edge cases that might occur.
\end{enumerate}

\subsection{Missing input validation in the legacy codebase}\label{sec:cust-attack-2}

This type of attack exploits legacy codebase security flaws that can appear when a system is updated. For example, when a system adds a new feature, it can still reach the same result using the old code or a mix of old and new code. 
In the case of a custodian, the expected result is to obtain the emission of a valid event on the source chain, to be relayed by the communicator to the destination chain. 
Let us consider that mandatory checks must pass to perform a given action. 
A vulnerability appears if an execution path to perform that action exists, such that at least one check can be bypassed. 
Note that adding a new way to perform an action to a system while keeping the old way active increases the number of execution paths. That number can get even bigger if it is possible to reach the same result using a mix of old and new code.

One approach to mitigate that risk is to limit the number of execution paths to perform an action, ideally to a single occurrence. 
Also, the developers should consider if the initial way of performing the action should be disabled or not.

In the next section, we will review 2 real-world examples of such attack vector.

\subsubsection{Real World Example}

\begin{figure*}
    \centering
    \includegraphics[scale=0.6]{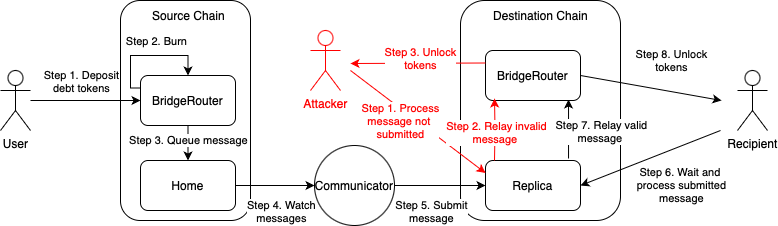}
    \caption{Overview of the attack of the Nomad Bridge.}
    \label{fig:nomad-diagram}
\end{figure*}

The first real-world example is the Nomad Bridge Hack that occurred in August 2022 and led to a loss of around \$190 million USD \cite{Zellic-nomad, samczsun-nomad}. 
Figure \ref{fig:nomad-diagram} shows this attack's general approach, which is explained in detail below. 

To understand the attack, the regular operation of the bridge should first be reviewed. During the routine bridge operation, tokens are burned via the \texttt{BridgeRouter} contract, which calls the \texttt{Home} contract to queue the newly generated message. The bridge maintains an internal record as a Merkle tree, which includes valid messages. The \texttt{Home} contract updates the Merkle root as new messages are received. The communicator relays the new Merle root to the \texttt{Replica} contract on the destination chain via the \texttt{update()} function. After a certain time window, where this Merkle root can be disputed for being incorrect, it can be used on the destination \texttt{Replica} contract. 
Then the user calls the \texttt{Replica} contract via \texttt{proveAndProcess()} with a proof of deposit on the source chain that is verified against the new Merkle root via \texttt{acceptableRoot()} and a message that is intended for the destination chain. Funds are released after successful execution. The proof is also stored along with the message so that it cannot be replayed at a later stage.

To simplify storage patterns and deployment of future contracts, the team implemented a feature that allows initializing a contract at a specific Merkle root from the \texttt{Home} contract so that all the previous messages do not need to be processed again by setting  \texttt{confirmAt[\_root] = 1}. 
The issue occurred when both \texttt{Home} and \texttt{Replica} contracts were freshly deployed, meaning the \texttt{Home} contract root was \texttt{0}. This led to a root of \texttt{0} being approved as valid, hence \texttt{confirmAt[0] = 1}.

During the attack, the \texttt{process()} function has been called with an invalid message, which is called \texttt{acceptableRoot(messages[invalid\_message])}. As the message did not exist, this call was interpreted as \texttt{acceptableRoot(0)}, as a non-existent map entry in Solidity\footnote{Solidity \cite{solidity} is a common smart contract language for EVM based blockchains.} defaults to 0. The \texttt{acceptableRoot()} function in itself does a check against \texttt{confirmAt[\_root]}, and as \texttt{confirmAt[0] = 1}, this function did execute successfully. During normal operation, \texttt{confirmAt[0]} should have been rejected. In brief, any invalid message can be successfully executed on the destination chain \texttt{Home} contract. 

The attack created a fraudulent message that claimed a burn of tokens on the source chain and requested a release of tokens on the destination chain. 

This vulnerability was severe enough that even unsophisticated attackers could exploit it easily by simply changing the recipient address in the transaction data. While many attackers copied transactions stealing 100 WETH or 1M USDC, the bug had the potential for much more significant exploitation. 
Skilled attackers could forge messages to steal all liquidity of an asset in a single call, making it a critical vulnerability.

\subsubsection{Solution}
When updating a system, it is important to consider the meaning of default values before and after the operation, to make sure that no assumption is broken. Also, the impact of the changes should carefully be assessed, especially for the main use case of the system and its edge cases.

\subsubsection{Real World Example}

\begin{figure*}
    \centering
    \includegraphics[scale=0.5]{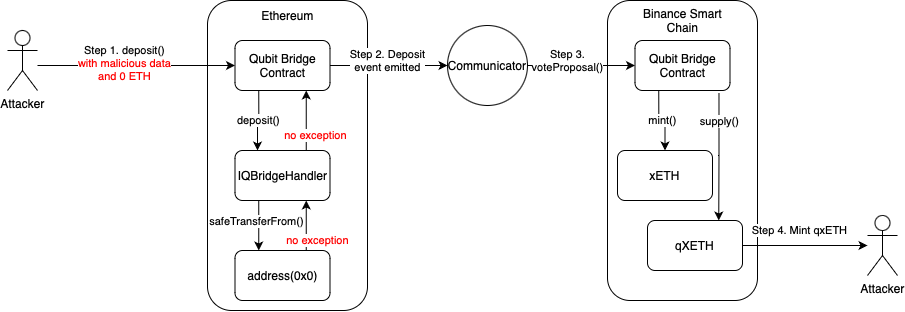}
    \caption{Overview of the attack of the Qubit Finance Bridge.}
    \label{fig:qubit-diagram}
\end{figure*}

The next real-world example is the Qubit Finance protocol hack that happened in January 2022 and led to a loss of around \$80 million USD \cite{qubit_certik, qubit_halborn, qubit_merkle, qubit_self}. 
Figure \ref{fig:qubit-diagram} shows this attack's general approach, which is explained in detail below.

Qubit Finance is a decentralized lending and borrowing platform. 
Qubit Finance manages a cross-chain bridge called QBridge that allows for swapping tokens between Ethereum and Binance Smart Chain.

The core issue for this hack is that legacy functionality for ETH deposit in the \texttt{deposit()} function was left in the contract when the new \texttt{depositETH()} function was added to handle ETH deposits differently. 

The attack consists of a sequence of cross-chain transactions, each following the same pattern described below.

The attacker called the function \texttt{deposit()} of the contract \texttt{QBridge} deployed on Ethereum with a custom malicious input without adding ETH to the transaction. The attacker used the value \texttt{0x0} for the deposited token address, representing the native token ETH.
It triggered a call to the function \texttt{deposit()} of the contract \texttt{IQBridgeHandler} that carries out verifications. The function checked that the address \texttt{0x0} was whitelisted, which was \texttt{true}, and that the amount was larger than the minimum amount, which was also \texttt{true}.
The last step was to invoke the function \texttt{safeTransferFrom()} to transfer the \texttt{0x0} tokens from the depositor's address, which also successfully returned \texttt{true}, resulting in the emission of a \texttt{Deposit} event. 
The communicator (a set of relayers) relayed that information to the Binance Smart Chain by interacting with the contract \texttt{QBridge}. 
As a result, an equivalent amount of xETH was minted as a BEP-20 \cite{bnb-chain_bep_20}, and the attacker received the corresponding amount of Qubit xETH. 
That token represents a proof of deposited xETH in the protocol Qubit Finance that can be later used as collateral to borrow assets from the protocol. 
 
Two main issues made that hack possible. The first issue was that the function \texttt{deposit()} should not have allowed deposits of native tokens (address \texttt{0x0}), as deposits of ETH were handled by the dedicated function \texttt{depositETH()}, and should have reverted instead. This allowed for the \texttt{0x0} address to be treated as an ERC-20 token rather than a native token. Also, as the function was expecting ERC-20 tokens, the amount of ETH attached to the transaction should have been verified.

The second issue was the execution of a non-standard function \texttt{safeTransferFrom()} at the token address \texttt{0x0}. The address \texttt{0x0} is an Externally Owned Account (EOA), which means that there is no code to run at that address. Also, any call to an EOA with the low-level instruction \texttt{call()} will not revert, which can be considered a successful execution. In the case of that hack, the function \texttt{safeTransferFrom()} was not checking that the address used was not an EOA. Hence, the contract did not revert because it considered the transfer successful. 

\subsubsection{Solution}

Legacy code and features increase the complexity of a system and require careful planning and testing to ensure unwanted behavior has not been added when updating the system, as seen in this attack. 
Also, the protocol used a modified version of the function \texttt{safeTransferFrom()} instead of the standard one from the library \texttt{SafeERC20} maintained by OpenZeppelin \cite{OpenZeppelin}. That modified method removed the verification that the target address must be a contract. Hence, the attack would have reverted by using the standard function.

\subsection{Compromised private keys}

In most cases, smart contracts have an admin or owner. This is an address with elevated privileges usually controlled by the developers or the protocol community. It may be an EOA or a smart contract.

\begin{figure}
    \centering
    \includegraphics[scale=0.5]{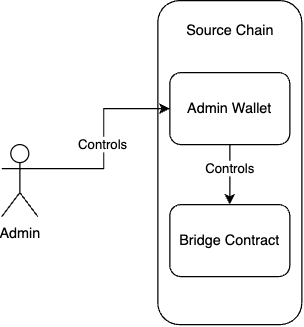}
    \caption{General example of an admin controlled custodian contract, commonly used in a centralised trusted bridge.}
    \label{fig:wintermute-general}
\end{figure}

In the case of bridge architecture, a common setup can be seen in Figure \ref{fig:wintermute-general}, where the bridge custodian contract has an admin with elevated permissions. This opens up multiple avenues for attacks. 

When the admin wallet's private key is compromised, an attacker can access the custodian bridge contract fully. The impact of this attack is limited or elevated by the extent of admin privileges, as the more significant the privileges, the more impact the attack has.

\subsubsection{Real World Example}

An example of such an attack in the wild is the Wintermute hack in June 2022 that led to around \$160 million USD loss to the company \cite{wintermute-rekt, halborn-wintermute}. 
The system is not exactly a bridge but has similarities with a centralized trusted bridge where only one entity can relay cross-chain messages. 
Still, this hack illustrates the impact of a compromised private key. 
Wintermute's admin wallet was compromised as the private key was generated with a tool that had a vulnerability in it.

Wintermute's admin address is a vanity address that was generated with a special tool called Profanity \cite{profanity-github}. A vanity address is a particular address which contains certain alphanumeric characters. Here, the Wintermute team decided to use a vanity address starting with several 0s because it decreases transaction gas costs.

It is revealed that Profanity has a critical bug, which was disclosed by 1inch \cite{1inch} a few days before the attack.
The issue with the Profanity tool lies in generating random private keys. The tool uses \texttt{std::random\_device} to generate a 32-bit seed, which is then fed into a pseudo-random number generator (\texttt{mt19937\_64}) to generate the 256-bit private key. However, this process is deterministic, meaning that you can predict the private key if you know the seed. Since there are only $2^{32}$ possible initial key pairs and the iteration process is reversible, it is possible to crack the private key from any public key generated by Profanity. This vulnerability compromises the security of the generated vanity addresses. It was initially assumed that it would take quite some time to brute force this according to 1inch, but a team at Amber group made a proof of concept where this process can be optimized to be computationally viable in reasonable time \cite{Amber_Group}.

Around the time of the bug disclosure, Wintermute removed all ether from their admin address. This suggests that they realized the address was vulnerable. However, they forgot to remove the address from their vault admins and only admins were allowed to withdraw funds, leaving the vault vulnerable to exploitation. With control over the compromised admin address, the attacker funded it with ether and used it to carry out the heist. It initiated transfers from Wintermute's vault contract, effectively stealing the tokens. It is important to note here that the contracts behaved as expected but were called by a compromised address.

\subsubsection{Solution}

Multiple mitigation strategies can be applied to avoid similar vulnerabilities. The first one would be avoiding the usage of public tools like Profanity, as without a thorough examination of the code it could contain unexpected issues. Secondly, private keys should be treated with the highest levels of security, and the bridge contract should have had a multi-signature (multi-sig) wallet as the admin account. Multi-sig wallets require multiple private keys to authorize transactions, so the attacker would have had to compromise multiple private keys to approve the transactions.
\section{Communicator Attacks}\label{sec:communicator}

In this section, we review two exploits targeting the communicator component of a bridge.
The first exploit uses compromised private keys to approve fraudulent deposits and withdrawals, while the second uses incorrect voting inside the communicator blockchain to approve fraudulent events.

\subsection{Compromised private keys of communicator}

\begin{figure}
    \centering
    \includegraphics[scale=0.6]{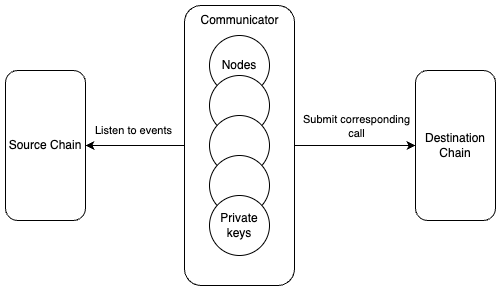}
    \caption{General implementation of a bridge with a five node communicator.}
    \label{fig:comm-general}
\end{figure}

The first type of attack covers two real-world examples when the private keys of the communicator get compromised. 
One of the standard design patterns in a trusted bridge is to have multiple validators that monitor for events on the source chain and submit the corresponding transaction on the destination chain. As seen in Figure \ref{fig:comm-general}, the communicator comprises five nodes with separate private keys. In the case of Figure \ref{fig:comm-general}, the policy could be that at least two out of five nodes must submit the event to the destination chain. Hence at least two nodes have to agree on the correctness of the deposit. 

This leads to multiple potential problems as security depends on the operators of those nodes and the security of those private keys. In the following two real-world examples, these attack vectors will be explored.

\subsubsection{Real World Example}

The first attack affected the Harmony Horizon Bridge in June 2022, which led to \$100 million USD being stolen from the project \cite{halborn-horizon, horizon_twitter}. The Horizon bridge enables the transfer of tokens from Ethereum, Binance Smart Chain, and Bitcoin to the Harmony network. This means that users can send various tokens such as ERC20, ERC721, BNB, BEP20 assets, and Bitcoin (BTC) to Harmony.

Attackers gained control of the multi-sig wallets used in the Horizon Bridge. The Horizon Bridge had a two out of five multi-sig setup, meaning two out of the five keys were needed to validate transactions. The attackers compromised two of these keys. 
Harmony Protocol confirmed that the attack was not due to vulnerabilities in their smart contract codes or the Horizon platform. The breach occurred because private keys were compromised, leading to the exploitation of the Horizon Bridge on the Ethereum network.
Harmony encrypted and stored the private keys, and a key management service was implemented. No single machine had access to multiple keys in plaintext. 
However, the attackers were able to access and decrypt several keys, including those used for unauthorized transactions involving BUSB, USDC, ETH, and WBTC.
Since the incident, the Harmony Protocol team upgraded the Ethereum side of the Horizon bridge to a four out of five multi-sig setup \cite{horizon_rugdoc}.

\subsubsection{Solution}

This incident highlighted the importance of traditional (or web2 security) in the web3 ecosystem. A system is as secure as its weakest link. Here it can be observed that it was not a smart contract compromise but a classic attack on the infrastructure to obtain the private keys. The system behaved as expected, as it was assumed that the holder of two out of five keys would not be malicious.

There are important takeaways from this attack: the multi-sig threshold of two out of five signatures is not always robust enough and even keys encrypted and stored with key management software can be compromised. Also, earlier in the paper, the example of the Wintermute hack highlighted that the way to generate keys can also be compromised.

\subsubsection{Real World Example 2}

This incident followed a similar pattern to the Harmony Horizon Bridge Hack. This attack occurred on March 23, 2022, affecting the Axie Infinitie's Ronin Bridge, but was only discovered on March 29, 2022 \cite{ronin-rekt, merkle-ronin}.

Axie Infinity is a blockchain-based game where players collect and breed digital creatures called ``Axies''.
Axies are stored on-chain as non-fungible tokens (NFTs). 
Players deposit ETH or USDC to Ronin, a sidechain specifically built for Axie Infinity, to purchase NFTs or in-game currency.
Ronin Network is a sidechain linked to the Ethereum blockchain. 
The Ronin bridge facilitates communication and asset interchangeability between Ethereum and Ronin.

The attacker compromised the Ronin bridge's pool of funds by compromising validator nodes operated by Sky Mavis and Axie DAO. The attacker drained 173,600 ETH and 25.5M USDC in two transactions from the Ronin bridge. The attacker used hacked private keys to forge fake withdrawals in two specific transactions, transferring 173,600 Wrapped Ethereum (WETH) and 25.5M USDC. 
At the time of the attack, the price of 1 WETH was \$3032.98 USD, which gives a total amount of \$526M USD worth of WETH. 
In Ronin, only five out of nine validator signatures were needed to recognize deposit and withdrawal events. The attacker gained control of five validator private keys, including four Sky Mavis validators and one Axie DAO validator.

The fact that Sky Mavis controlled four out of nine validators and the attacker managed to compromise those keys raises concerns about weak security practices. No single entity should operate a significant number of nodes alone.

The attacker found a backdoor through Ronin's gas-free RPC node \cite{ronin-rekt}, which they abused to get the signature for the Axie DAO validator. 
Sky Mavis had previously requested permission for Axie DAO to authorize transactions on its behalf, but the permissions were never revoked. Once the attacker accessed Sky Mavis systems, they obtained the Axie DAO validator's signature using the gas-free RPC node\footnote{A RPC (``Remote Procedure Call'') node is a type of computer server that allows users to read data on the blockchain and send transactions to different networks.}.

\subsubsection{Solution}

Multiple mitigation strategies could be adopted to reduce the chances of such attack occurring such as: 

\begin{enumerate}
    \item Monitoring. A concrete monitoring system should be in place so that this attack could have been detected straight away and not multiple days after.
    \item Decentralisation. One entity should not hold such a significant proportion of power.
    \item Rigor. Care should be paid to both on-chain and off-chain components of the system, because a system is as strong as its weakest link.
\end{enumerate}

\subsection{Unexpected validator behavior in the off-chain communicator}

Another approach to building a communicator network is decentralized, where the off-chain component is not run by one entity but rather by a range of entities. 
One way of organizing these entities is to have another blockchain for the communicator.

\subsubsection{Real World Example}

\begin{figure*}
    \centering
    \includegraphics[scale=0.6]{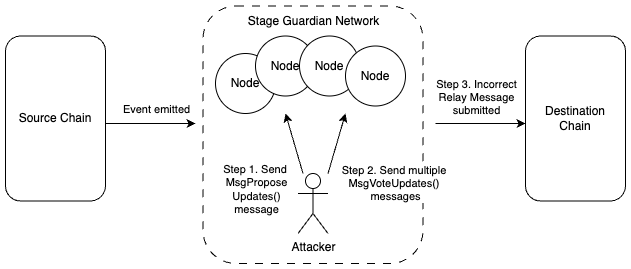}
    \caption{Example of behavior of the Celer cross-chain protocol where the validator nodes vote on the events to be processed.}
    \label{fig:comm-celer}
\end{figure*}

On May 24, 2023, a vulnerability was disclosed by Jump Crypto, which was discovered in the communicator part of the Celer cross-chain protocol \cite{jump-celer}. 
Figure \ref{fig:comm-celer} outlines the key steps taken to exploit the vulnerability, while the rest of this section details the exploit. 
If found in the wild, this vulnerability could have led to a loss of \$30 million USD.

Celer built its bridge and cross-chain communication products on the Stage Guardian Network. 
This network is a proof-of-stake blockchain based on Cosmos (see, e.g., \cite{DBLP:conf/icpads/WuLWLZ22}).
As part of this design, validators act as the communicator and monitor the Celer contracts for transfers and messages in order to forward them to the destination chain.

Users who want to bridge a token call the \texttt{send()} method of the Celer bridge contract, which locks the tokens and emits a \texttt{Send} event. 
The event is observed by a sensor node (the communicator), responsible for monitoring supported chains for event logs. 
The sensor node combines the events received and sends a \texttt{MsgProposeUpdates} message containing the discovered events to other nodes.
As a protection against malicious updates, Celer relies on a voting mechanism where SGN nodes verify proposed updates. Each node votes on the outcome of a proposed update by sending a \texttt{MsgVoteUpdates} message with ``yes'' or ``no'' votes.

At the end of the block, the function sums the ``yes'' votes weighted by the appropriate stake and applies the update if more than 2/3 of the total staked amount voted ``yes''. 
However, the code lacks a check to prevent a validator from voting multiple times on the same update, allowing them to multiply their voting power.
Without this check, a malicious validator can propose and approve invalid or malicious updates. 
For example, they could forge a large token transfer and trigger the execution of the Relay message on the destination chain, transferring tokens to their account.

Celer has measures to mitigate the impact of such attacks, including delaying outgoing transfers and using a volume control mechanism. 
The contracts can also be paused. 
These measures were effective.

\subsubsection{Solution}

This vulnerability disclosure highlighted the importance of scrutinizing not just the on-chain component but also the off-chain parts of the system. The issue was discovered as soon as the codebase for the communicator was open-sourced. However, this issue was not covered by the bug bounty program that Celer runs, lacking an incentive for security researchers to investigate.

Hence a mitigation strategy is closer attention to off-chain components and other aspects of the system that might be leveraged to attack the protocol. 
A system is as strong as its weakest component.

\section{Related and Future Work}\label{sec:related}

This paper continues the line of work initiated by Lee et al.~\cite{prev_paper_QS}. That paper set out a general bridge architecture design and identified common risk areas via analysis of documented bridge hacks. 
This paper took the same approach but focused on attacks that occurred from December 2022 until May 2023. 
There also has been work conducted related to bridge security that will be reviewed in this section. 

The area of cross-chain communication has been extensively studied and researched. Many papers have done overviews of cross-chain solutions such as Belchior et al.~\cite{interoperability_2023} which proposes a framework to compare interoperability mechanisms, Ou et al.~\cite{OU2022109378} which proposes an overview of the existing cross-chain technologies and the main challenges, and Zamyatin et al.~\cite{sok_cross_chain} which propose a framework to design new and evaluate existing cross-chain communication protocols.

A significant amount of work has also been conducted on detection and prevention methodologies. Zhang et al.~\cite{xscope_2023} propose a classification to describe bridge attacks: Unrestricted Deposit Emitting (UDE), Inconsistent Event Parsing (IEP) and Unauthorized Unlocking (UU). The paper then derives logic rules to detect these attacks and concludes with a tool called XScope that is based on the previously derived rules. Belchior et al.~\cite{Hephaestus} propose a system called Hephaestus, that extends Hyperledger Cactus \cite{cactus}. This system generates cross-chain models from local transactions to allow for monitoring of the application, hence identifying malicious or unexpected behaviour.

Abebe et al.~\cite{cross_chain_framework} have introduced a cross-chain communication risk framework. It has been built based on previous bridge hacks and non-exploited vulnerability disclosures. Our work is complementary to theirs and could help to inform future versions of their framework, as it analyses the hacks in detail and classifies them. The initiative of Kiepuszewski et al.~\cite{l2_beat_framework} should also be highlighted where the authors propose a risk framework for assessing the security profile of different bridge architectures.

For future work, we will continue our efforts to study the latest hacks of bridges to ensure that they are not repeated.
It would also be interesting to further explore mitigation strategies and see how those would have affected the identified hacks. 
More generally, there is a need for more rigor in standardized security best practices throughout the lifecycle of a bridge system: design, development, deployment, operations, incident response, upgrades and deprecation.
\section{Conclusion}\label{sec:conclusion}

In conclusion, we have provided a comprehensive analysis of recent cross-chain bridge hacks, shedding light on the vulnerabilities and weaknesses exploited in these incidents. The findings emphasize the urgent need for enhanced security measures and standardized best practices in the development and operation of cross-chain bridges. The substantial financial losses incurred by these attacks highlight the critical importance of addressing the vulnerabilities in bridge architecture and communication protocols. By understanding the nature of these attacks, potential countermeasures can be proposed to mitigate risks and enhance the security of cross-chain bridges. Moreover, this research contributes to the broader understanding of the challenges the blockchain industry faces in achieving seamless interoperability. Developing industry-wide standards and continued research efforts are crucial to ensure the integrity and resilience of cross-chain bridges.

\paragraph*{Acknowledgements}

The authors would like to thank especially their colleagues Sebastian Banescu, Mohsen Ahmadvand, and Marius Guggenmos for their insights and fruitful inputs that were incorporated in this paper.

\bibliographystyle{unsrt}
\bibliography{references}


\end{document}